\documentclass[fleqn,10pt]{wlscirep}
\PassOptionsToPackage{normalem}{ulem}
\usepackage{ulem}

\providecolor{added}{rgb}{0,0,1}
\providecolor{deleted}{rgb}{1,0,0}

\usepackage[utf8]{inputenc}
\usepackage[T1]{fontenc}
\title{Randomness assisted in-line holography with deep learning}

\author[1]{Manisha}
\author[1,2]{Aditya Chandra Mandal}
\author[1]{Mohit Rathor}
\author[3]{Zeev Zalevsky}
\author[1,*]{Rakesh Kumar Singh}
\affil[1]{Laboratory of Information Photonics and Optical Metrology, Department of Physics, Indian Institute of Technology (Banaras Hindu University), Varanasi, 221005, Uttar Pradesh, India}
\affil[2]{Department of Mining Engineering, Indian Institute of Technology (Banaras Hindu University), Varanasi, 221005, Uttar Pradesh, India}
\affil[3]{Bar-Ilan University, Faculty of Engineering and Nano Technology Center, Ramat-Gan, Israel}

\affil[*]{krakeshsingh.phy@iitbhu.ac.in}

\usepackage{subfig}
\usepackage{float}
\usepackage{graphicx}
\usepackage[autostyle]{csquotes}

\begin{abstract}
We propose and demonstrate a holographic imaging scheme exploiting random illuminations for recording hologram and then applying numerical reconstruction and twin removal. We use an in-line holographic geometry to record the hologram in terms of the second-order correlation and apply the numerical approach to reconstruct the recorded hologram. 
The twin image issue of the in-line holographic scheme is resolved by an unsupervised deep learning(DL) based method using an auto-encoder scheme. This strategy helps to reconstruct high-quality quantitative images in comparison to the conventional holography where the hologram is recorded in the intensity rather than the second-order intensity correlation. Experimental results are presented for two objects and a comparison of the reconstruction quality is given between the conventional inline holography and the one obtained with the proposed technique. 
\end{abstract}
\begin{document}

\flushbottom
\maketitle
%
%
\thispagestyle{empty}

\section*{Introduction}
Digital holography(DH) has  emerged as a powerful tool for recording and reconstructing the amplitude and phase information of the wave \cite{r0,tahara2018digital,javidi2021roadmap,Rosen_2023}. The ability of the DH to retrieve complex amplitude information has a wide range of applications in 3D displays \cite{park2019ultrathin}, microscopy\cite{park2009speckle}, biomedical imaging \cite{kemper2008digital}, and many more. The DH provides a spatially resolved quantitative phase images and depth reconstructions. In-line, off-axis, and phase shifting are a few widely used schemes. In an off-axis holography, two coherent and angularly separated beams interfere to record the hologram information\cite{leith1962reconstructed}. Since available digital detectors have a limited pixel pitch, angular separation between interfering beams introduce a limitation for the off-axis DH. Moreover presence of unmodulated term and conjugate limits utilization of the full bandwidth in an off-axis DH geometry. Phase shifting is another technique that uses multiple recordings of the same object with phase shifts in the reference wave \cite{zhang1998three}.
Among several holographic techniques, in-line holography has a compact design, and a high space bandwidth product(SBP)  \cite{kim2010principles,gabor1948new}. In-line holography schemes can 
 be designed by using a single path and obtained by the interference of diffracted and un-diffracted waves emerging from the object \cite{gabor1948new}. However, a bottleneck of in-line holography is the ubiquitous twin image problem. Various techniques have been developed to resolve this issue using optical and computational methods\cite{latychevskaia2007solution,zhang2018twin,li2020deep}.

A quality of reconstruction in the DH depends on the recording configurations. Due to digital recording and constraints on the detectors, improving resolution is an emerging interest in the DH. Resolution in a digital holographic setup is influenced by factors such as; the numerical aperture, detector pitch, and diffraction. In past, various techniques have been proposed to improve the resolution in the DH and some of these techniques are lowering the wavelength \cite{faridian2010nanoscale}, down-sampling the detector pitch \cite{song2016sparsity}, increasing the effective numerical aperture \cite{jiang2015high,mico2006superresolution}, expanding computational bandwidth \cite{zhang2020resolution}. Recently several new ideas have been proposed to enhance the SBP \cite{li2015space,park2009speckle,yilmaz2015speckle}. Structured light illumination has also been used to improve the image quality and resolution \cite{gustafsson2000surpassing,gustafsson2005nonlinear,heintzmann1999laterally,heintzmann2017super,gao2013structured,chowdhury2017refractive,zheng2014digital}. Zheng et al used structured illumination in different orientations combined with an iterative algorithm to enhance the spatial resolution in the DH \cite{zheng2014digital}. Speckle field illumination has also been used in the DH for high-resolution imaging and in enlarging field of view \cite{park2009speckle,zheng2015autofocusing,garcia2005synthetic,vinu2019speckle,meitav2016point,bianco2012clear,funamizu2019enhancement}.
However, these speckle illumination methods need recording of several holograms with the random illumination patterns for proper cancellation of randomness.


On the other hand, use of the  randomness rather than cancellation has shown significant potential in imaging such as ghost imaging, ghost diffraction microscopy \cite{Manisha2023ghost,vinu2020ghost}, resolution enhanced wide field imaging \cite{yilmaz2015speckle} and many more.
 Memory effect \cite{freund1988memory} within a speckle pattern offers a high-resolution imaging systems\cite{van2011scattering}. Correlations have been exploited in super-resolution optical fluctuation imaging (SOFI) with dynamics near field speckle patterns. \cite{choi2022wide}.It has been demonstrated that higher-order (n) intensity correlations improve the resolution by a factor of  $\sqrt{n}$ \cite{li2019beyond} and such  higher-order correlations have been utilized for improving resolution beyond the diffraction limit \cite{li2019beyond,liu2019resolution}. A sub-Rayleigh imaging has been demonstrated using a second-order correlation measurement\cite{oh2013sub}.
However, majority of these correlation techniques mainly deal with amplitude object, without any phase signature of the signal except a recent work \cite{dou2023sub}.In the reference \cite{dou2023sub}, a sub-Rayleigh dark-field imaging by speckle illumination is demonstrated and an autocorrelation image is presented for a binary phase object.

In this paper, we propose and demonstrate a new method to record an in-line hologram in intensity correlation rather than the intensity as in conventional holography. This is aimed to enhance the quality of reconstruction in the holography and demonstrate it by reconstruction of a complex-valued object in the inline DH. A common issue of twin images in the in-line holography is tackled with a deep learning method by utilizing an auto-encoder model.

We use a deep learning approach for a single-shot image reconstruction without a large training dataset \cite{li2020deep}. The auto-encoder minimizes a well-defined objective function to reduce noise and remove twin image instead of suppressing it. Our results show that a neural network equipped with the proposed correlation methods provides an enhanced image quality of the complex-valued object. The proposed technique is experimentally verified and results are presented for two different cases; i.e. a conventional holographic recording and hologram recording by second-order correlation. A comparison of conventional holography and the proposed method highlight high-quality reconstruction in the new technique. A theoretical background and experimental demonstrations are discussed below.

\section*{Theory and Methodology}
A conventional recording of an in-line hologram and its comparison with proposed technique is shown in Fig. 1. Diffracted and undiffracted beams from the object interfere at plane 1 and make an in-line hologram as a distribution of the intensity. This inline hologram is imaged and digitally recorded at plane 2 as shown in Fig. 1(a). An aperture is placed in the imaging system to control the numerical aperture and consequently analyze its impact on the image quality.
An intensity of the in-line hologram at plane 1 is given by,
\begin{equation}
I_H(\rho)=\left|E_H(\rho)\right)|^2=\left|E_D(\rho)\right|^2+\left|E_d(\rho)\right|^2
+E_D^*(\rho) E_d(\rho)+E_D(\rho) E_d^*(\rho)
\end{equation}
where $E_H(\rho)$=$E_{D}(\rho)$ + $E_{d}(\rho)$ is the complex field at plane 1.  $E_{D}(\rho)$ and $E_{d}(\rho)$ are diffracted and un-diffracted beams respectively and $\rho$ is the spatial coordinate at plane 1. The optical field at plane 2 is represented as,
\begin{equation}
E(r)=\int E_H(\rho))  h(r-\rho) d \rho
\end{equation}
 where the digitally recorded hologram is $I=|E_H \circledast h|^2$,
$\circledast$ represents the convolution operator and $r$ is the spatial coordinate at plane 2. $h(r-\rho)$ represents the point spread function (PSF) of a diffraction-limited imaging system. The PSF for a diffraction-limited imaging lens is an airy disk and is represented as,

\begin{equation}
h\left({r}, {\rho}\right) \propto \int d {\rho}_0 P\left({\rho}_0\right) \exp\Bigg\{\frac{i \pi}{\lambda}\left[\frac{\left({\rho}_0-{r}\right)^2}{d_1}+\frac{\left({r}-{\rho}_0\right)^2}{d_2}-\frac{{\rho}_0^2}{f}\right]\Bigg\}
\end{equation}

 where P($\rho_0$) is the pupil function describing the effective entrance pupil,
and $d_1$ and $d_2$  are the distances from the lens to
the object plane and image plane, respectively, and $d_1$=$d_2$=2f.
Eq. 2 represents a digital recording of the hologram which is numerically reconstructed to recover the complex-valued object from the DH.

Now, consider a scatterer at the plane 1 which hides the direct recording of the hologram from the detector. A random scatterer in the path of the light scrambles the wavefront and generates a speckle pattern as shown in Fig. 1(b).
 A single realization of the field immediately after the scatterer is given by,
\begin{equation}
 E(\rho,t)=E_H(\rho,t) \exp \left[i \phi(\rho,t)\right]
\end{equation}
 
where $\phi(\rho,t)$ is the random phase introduced by the scatterer and $t$ represents time corresponding to different random patterns.
The complex field at the recording plane 2 is represented as,
\begin{equation}
E(r,t)=\int E(\rho,t))  h(r-\rho,t) d\rho
\end{equation}
\begin{figure}{}
\centering\includegraphics[width=9cm]{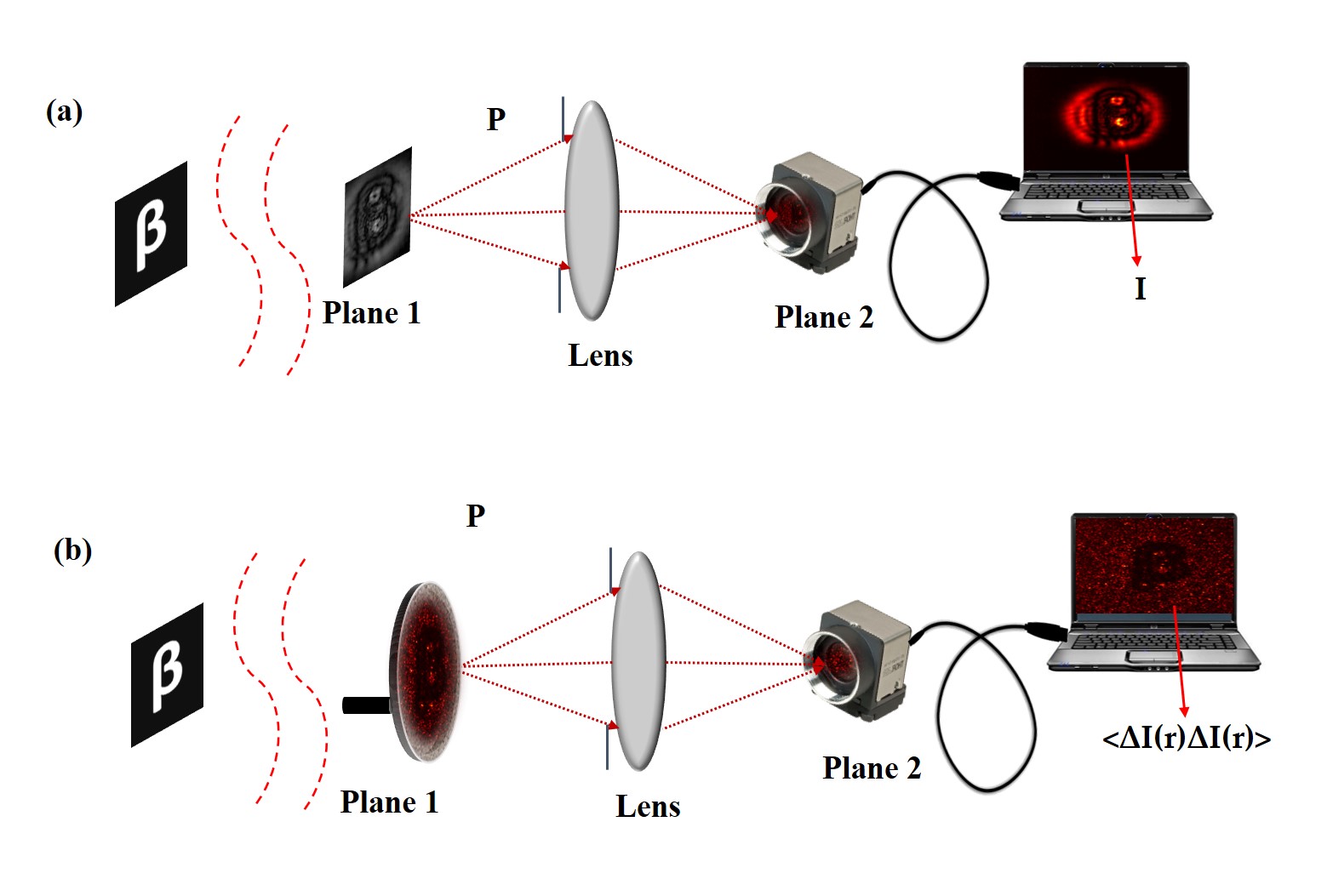}
\caption{A comparison of (a) conventional and (b) proposed technique}
\end{figure}

The random intensity at plane 2 is
\begin{equation}
    I(r,t) =\left|E(r,t)\right) |^2 
\end{equation}
Imaging of a hologram through a scatterer has been used in the past for looking around corners \cite{singh2014looking} and in the correlation holography \cite{kumar2014recovery}. These methods rely on recovering conventionally recorded hologram i.e, $I(\rho)$ from the averaged random fields.
On the other hand, here we propose and experimentally demonstrate a new technique by recording the hologram in terms of the second-order correlation of intensity rather than intensity for high-quality holographic reconstruction.

Following Eq. 6, intensity fluctuation is given as,
 \begin{equation}
 \Delta I(r,t)= I(r,t) - \langle I(r,t)\rangle
\end{equation}
 where angular bracket $\langle ...\rangle$ represents ensemble average and $\langle I\rangle$ is mean intensity.
 
We introduce and use correlation of intensity fluctuations to record the hologram and this correlation function is represented as, 
\begin{equation}
g^{2}(r,r)= \langle \Delta I(r,t)\Delta I(r,t)\rangle
\end{equation}

Eq. 8 represents our quantity of interest to record the in-line hologram. In comparison to the conventional hologram recording 
in I(r) , the proposed technique records hologram as $g^{2}(r,r)$.
For the Gaussian random field, second-order intensity correlation can be expressed as a modulus square of field correlations as,
 \begin{equation}
g^{2}(r,r)=|\langle E^{*}(r,t)E(r,t)\rangle|^2
\end{equation}
where asterisk $*$ represents the complex conjugate.
The second-order field correlation is represented as,
\begin{equation}
  \langle E^{*}(r,t)E(r,t)\rangle=\iint\ E^{*}(\rho_1)E(\rho_2)\langle exp (-i \phi(\rho_1,t))\exp \left(i \phi(\rho_2,t)\right\rangle_T\ h^{*}(r-\rho_1)h(r-\rho_2) d\rho_1 d\rho_2 
\end{equation}
For a spatially incoherent source, the correlation at plane 1 is represented as,
\begin{equation}
\langle\exp  i(( \phi(\rho_2,t)-\phi(\rho_1,t))\rangle_{T}=\delta(\rho_2-\rho_1)
\end{equation}

Substituting Eq. 11 in Eq. 10 leads to
\begin{equation}
\langle E^{*}(r,t)E(r,t)\rangle=I(\rho) \circledast h^{2}(r-\rho)
\end{equation}
Therefore second-order intensity correlation is expressed as,
\begin{equation}
    g^{2}(r,r)=\left|I \circledast h^{2}\right|^2
\end{equation}

For a uniform source, i.e, $I(\rho)=1$ ;the second order intensity correlation transforms to,
\begin{equation}
g^2\propto h^4
\end{equation}
Equation 14 represents that the  correlation of intensity fluctuations is proportional to the fourth power of the PSF of an imaging lens. A comparison of Eq. 2 and Eq. 14 shows that the recording quality is enhanced in the higher-order correlation due to a narrower size of the PSF as compared to the conventional  case. This is quantitatively analyzed and experimentally tested in the coming section. Although the quality of recording the inline hologram is improved in the randomness-assisted approach, reconstruction of an inline hologram demands the removal of the twin images. This is tackled by using an auto-encoder deep learning method as explained in the coming section.

\section*{Experimental recording}
\subsection*{Experimental recording and numerical reconstruction of inline hologram}

A monochromatic laser beam of wavelength 633nm (Thorlabs, Model No. HNL 150L) is spatially filtered and collimated by a spatial filter assembly, and a lens L1.
A beam splitter (BS) divides the incident beam into two equal reflected and transmitted components. The beam transmitted from BS is used to illuminate a reflective type spatial light modulator (SLM). This SLM is having 1280×768 pixels with a pixel pitch of 20µm. An object is inserted in the incident light using the SLM and the light reflected from the SLM is folded by the BS to propagate toward the plane 1(without RGG). Interference of diffracted and undiffracted waves from the object makes an inline hologram at the plane 1 and this plane is imaged by lens L2 at the camera plane. Focal length of lens L2 is 100mm and a variable aperture A2 is placed to control the numerical aperture of the imaging lens as shown in Fig. 2. A CMOS camera with 1024×1280 pixels, and pixel size 5.4µm, records the in-line hologram.
\begin{figure}{}
\centering\includegraphics[width=9cm]{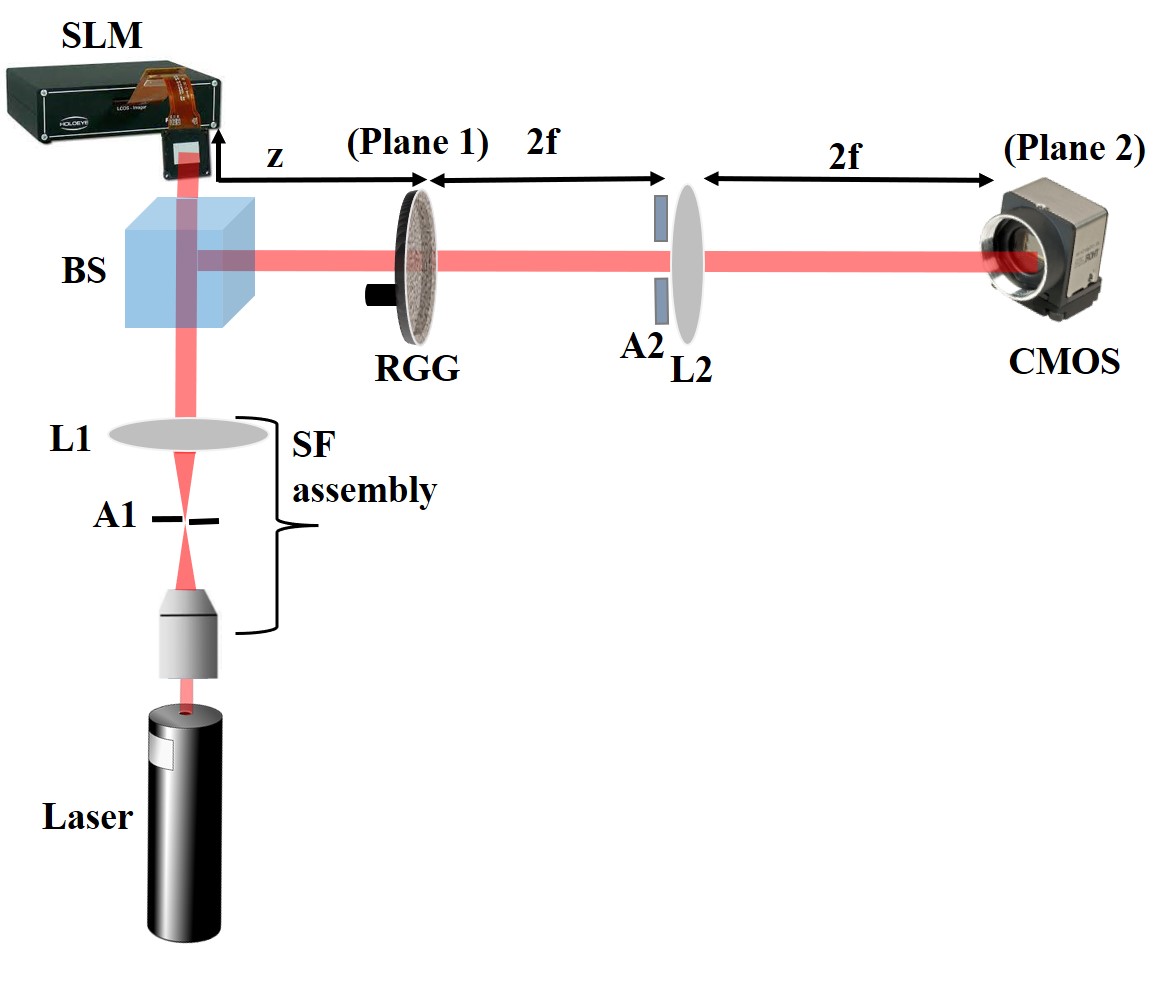}
\caption{Experimental Setup; SF: Spatial filter, L1 and L2:Lens with focal lengths 200 and 100mm respectively, BS: Beam splitter, SLM: Spatial light modulator, RGG: Rotating ground glass, A1 and A2:Apertures, CMOS Camera: Complementary metal oxide semiconductor camera}
\end{figure}

In order to replace conventional hologram recording by recording in the correlation of intensity fluctuations, we insert a rotating ground glass(RGG) at plane 1 in the experimental system. An inline hologram at plane 1 is now scrambled due to the random phase introduced by the RGG. Therefore recorded pattern at the camera plane represents a speckle pattern for a particular position of the RGG and the pattern is represented as $| E \circledast h|^2$. This random phase is common to both the diffracted and undiffracted beams at the plane 1, thus hologram information can be retrieved by averaging the random patterns in spite of randomness in the path of the light. The randomly scattered inline hologram is recorded using different orientations of the RGG at different times t.
For experimental demonstrations of the proposed technique, we used 150 frames which are independently generated by the RGG. A series of random patterns with independent randomness are utilized to measure the correlation of intensity fluctuations and to record the inline hologram. 
In the randomness-assisted recording of an inline hologram, a correlation of intensity fluctuations  is measured by estimating correlations of intensity fluctuations over varying random patterns. We considered averaging over 150 random patterns and is represented as
\begin{displaymath}\sum_{n=1}^{\ 150}\left  [ I_n(r) -\langle \ I(r) \right\rangle ]\left  [ I_n(r) -\langle \ I(r) \right\rangle ]\end{displaymath} 
This equation represents the correlation of the fluctuations of intensity patterns corresponding to 150 independent random patterns generated by the ground glass.
Each shot of the recorded image has a noisy pattern for fixed orientation of RGG at a time t. We make use of averaging over a number of random patterns and $g^2(r,r)$ is estimated.
This results in a hologram with reduced noise and an inline hologram appearing as a distribution of $g^2(r,r)$.
Experiment is performed for two different objects, a Greek letter $\beta$ and 
 an English letter "V" . An inline hologram of letter $\beta$ is shown in Fig. 3(a) recorded as a distribution of the intensity $I(r)$ and recording of an inline hologram of the same object as a distribution of second order intensity correlation $g^2(r,r)$  is represented in Fig. 3(b). The detailed procedure of retrieving quantitative information from the inline hologram with and without randomness is described below.

\subsection*{Deep learning for numerical reconstruction of inline hologram}
In the past, different methods have been proposed to tackle reconstruction of the in-line hologram, including the phase retrieval algorithms \cite{beleggia2004transport,denis2005twin,liu1987phase,fienup1982phase,gerchberg1972practical,latychevskaia2019iterative}. Accurate recovery of the phase can lead to strong suppression of the twin image \cite{flewett2012holographically,gnetto2022solving}.
We use a deep learning approach for reconstructing the in-line hologram using an encoder-decoder architecture \cite{li2020deep}. This network maps a high dimensional input image,  $\alpha$, into a low dimensional latent code, $\beta$. The mapping from the input image to the latent code and the reconstruction of the output from the latent code are represented by the functions $F_{\text{encode}}(\alpha) = \beta$ and $F_{\text{decode}}(\beta) = \hat{\alpha}$, respectively. For a supervised image restoration, the output image is compared to a ground truth image, and the error between them is penalized. This model can learn the uncorrupted and realistic features of the input image through an iterative process using an untrained network, allowing it to converge on a denoised output. To address the twin image issue in an inline hologram, a physical model is designed for training the auto-encoder. The auto-encoder, denoted as $F_{auto-encoder}(\alpha, w)$, is trained to minimize the error between the captured in-line hologram $H$ ($I(r)$ or $g^2(r,r)$ ) and the reconstructed in-line hologram, which is the forward-propagated result of the reconstructed complex object wave to the plane 1 using $F_{transmission}$. The weights $w$ are randomly initialized at the start of training. The goal is to find the weights $w$ that minimize the error between the captured in-line hologram and the reconstructed in-line hologram. The objective function can be written as:
\begin{equation}
w={\arg \min }|H-F_{transmission}(F_{auto-encoder}(H, w))|_2^2
\end{equation}
Where $| . |_{2}$ is $L_2$ norm. As depicted in Fig. 4, here we use an "Hourglass" type auto-encoder architecture. This encoder transfers the fixed network input into lower-dimensional space, encoding steps, and decoding steps are done by reconstructing the complex object from their representation by the latent variables. We use a wavelet transform and its inverse transform to provide downsampling (encoding) and upsampling (decoding), which substitute strided convolution, pooling, or interpolation\cite{li2020deep}.
The proposed method is shown step-by-step in Fig. 5. The in-line hologram obtained from Eq. 14 is first back-propagated to the object plane to determine the amplitude and phase distributions. These are then fed into the auto-encoder model. The network outputs the reconstructed amplitude and phase, which are then forward propagated to the in-line hologram plane using $F_{transmission}$ to get a reconstructed in-line hologram. The mean absolute error (MAE) loss between the original hologram and reconstructed hologram is calculated to update the model's weights."

\begin{figure}{}
\centering\includegraphics[width=9cm]{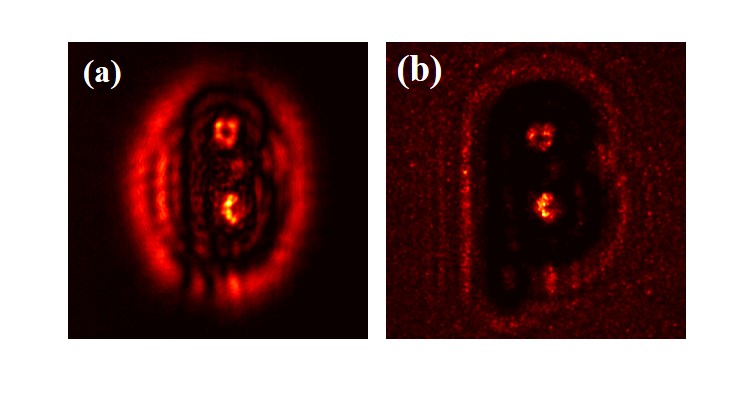}
\caption{(a)Conventional inline hologram for $\beta$ ;(b)Inline hologram in terms of second order correlation}
\end{figure}

\section*{Results and Discussion}
We record inline holograms in two different experimental configurations as shown in Fig. 1. In the first case, an inline hologram is recorded in free space i.e, without a scatterer. The inline hologram at plane 1 is imaged by a lens L2 with a variable aperture A2in front of the lens L2. 
In the second case, when a scatterer is placed at plane 1, a speckle is generated due to the randomness and interference of randomly scattered waves. Light is scrambled into the speckle pattern without any direct resemblance with the hologram I(r).
\begin{figure}{}
\centering
\includegraphics[width=16cm]{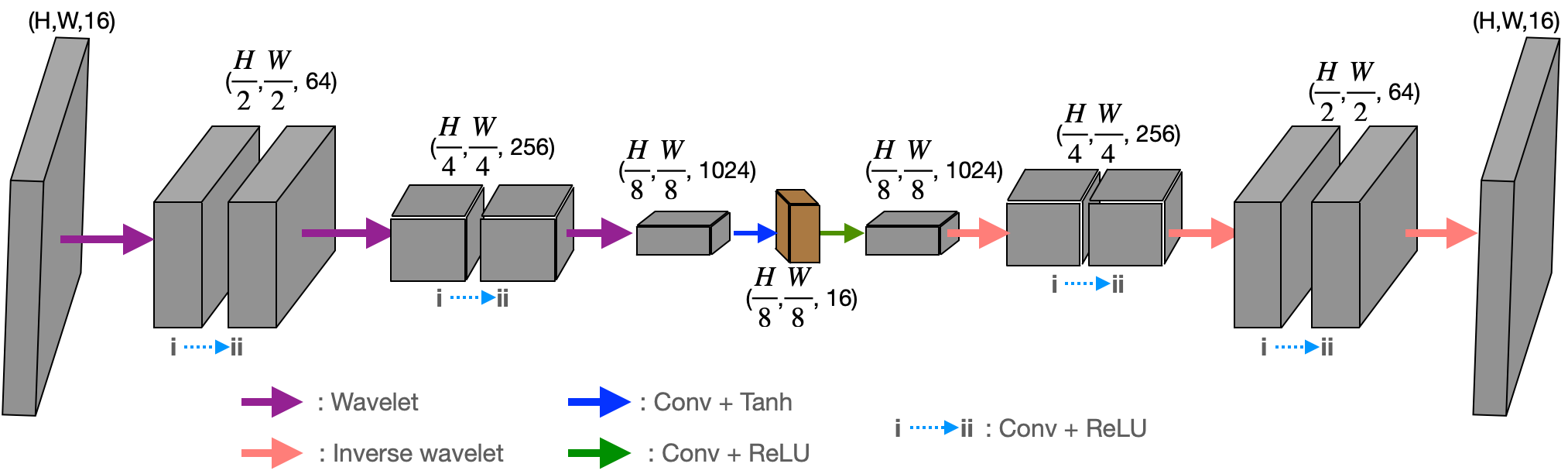}
\caption{An auto-encoder network utilizing deep convolutional layers in an hourglass architecture}
\end{figure}
For randomness-assisted inline hologram recording, a series of random patterns are captured and these random patterns are digitally processed to obtain the second-order correlation of intensity.
The inline hologram is back-propagated to the object plane by using an angular spectrum method. However,inline configuration of recording introduces a twin image in the reconstruction. Therefore a deep learning is applied to quantitatively reconstruct the inline hologram without a twin image. We employ the auto-encoder model to resolve this problem and this is implemented using the PyTorch framework and runs on a GPU workstation with an Nvidia tesla P100 graphics card. The Adam optimization algorithm is utilized with a fixed learning rate of 0.005. To train the network, we used angular spectrum back-propagation reconstruction of the in-line hologram as input and use the training process for 4000 iterations until we get the final reconstruction. A complete flow chart for the same is given in Fig. 5.
\begin{figure}{}
\centering\includegraphics[width=13cm]{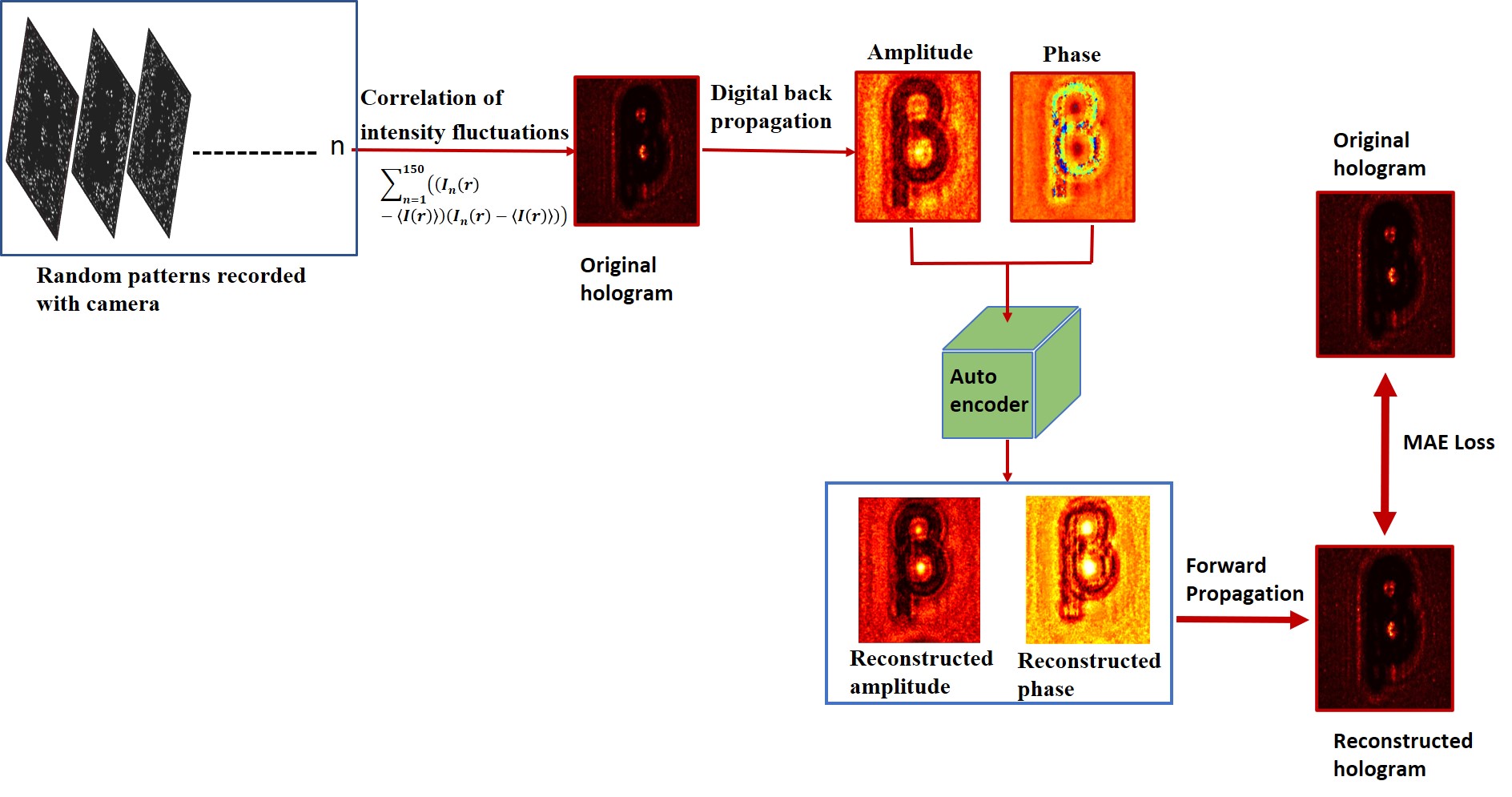}
\caption{A flow chart for the proposed scheme}
\end{figure}

Reconstruction results are presented for two different objects viz, $\beta (size:2.8mm \times 1.48mm)$ and letter $V(2.84mm\times 2.64mm)$ as shown in Fig.6 and Fig. 7 respectively in free space and randomness assisted geometry. Fig. 6 (a) and (b) are the amplitude and phase of letter $\beta$ in the inline holography without scatterer.Fig.6(c) and (d) represent the retrieved amplitude and phase for the randomness-assisted hologram recording. Fig. 7(a) and (b) are the amplitude and phase for letter V in the inline holography without scatterer and Fig. 7(c) and (d) show reconstructed amplitude and phase distribution for the proposed randomness-assisted inline holography technique. Comparison of inline holography in free space and through randomness shows that reconstruction quality is significantly enhanced in the hologram recording with the second-order intensity correlation. Edges and frequency contents are enhanced in the holography with $g^2(r,r)$ in comparison to the holography with $I(r)$. This is observed by accessing the spatial frequency contents of the reconstructed object for both inline holographic recording methods i.e, $I(r)$ and $g^2(r,r)$.
For a quantitative evaluation and comparison of the quality of reconstruction of both hologram recording methods, we also introduce two parameters, such as visibility and reconstruction efficiency \cite{hillman2013digital}. Visibility gives a measure of the degree to which the signal reconstruction can be differentiated from the background noise. It is defined as the ratio of the signal region's average intensity level to the background region's average intensity level in the reconstructed image. On the other hand, reconstruction efficiency is the proportion of the signal region's measured power to the total of the signal and background regions' measured powers in the reconstructed image. Visibility values for letter $\beta$ in the conventional and proposed scheme are 0.042 and 0.21, whereas respective reconstruction efficiency values are 0.55 and 0.885. For other object V, visibility and reconstruction efficiency values are 0.081 and 0.67 for conventional case. The improved Visibility and reconstruction efficiency values for object V  in the proposed scheme are  0.55 and 0.92.

\begin{figure}{}
\centering\includegraphics[width=10cm]{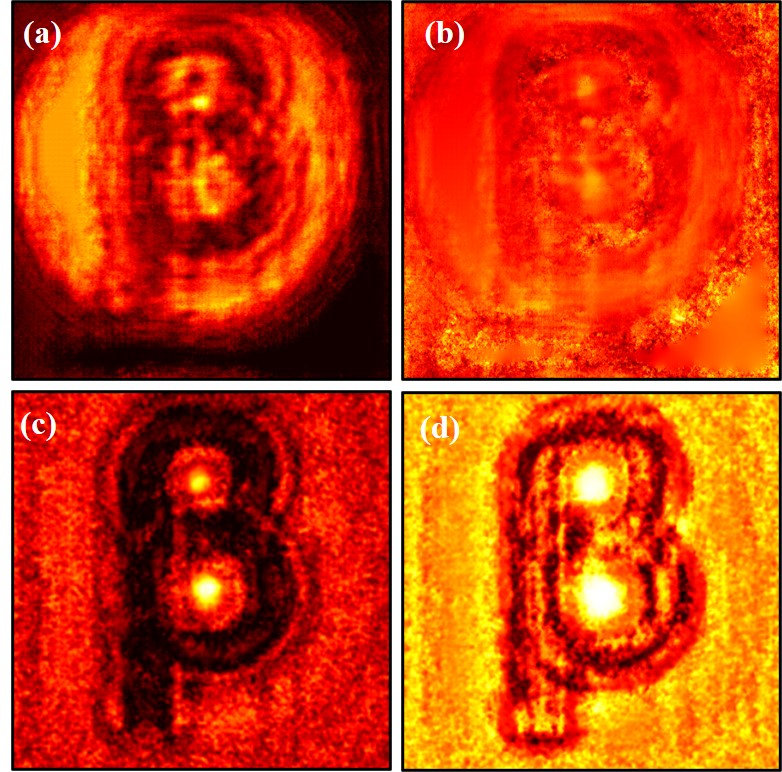}
\caption{(a-b)Conventional case results for letter 
$\beta$ ;(a)Amplitude (b) Phase,(c-d)Proposed scheme based results for letter $\beta$ ;(a)Amplitude (b) Phase}
\end{figure}

\begin{figure}{H}
\centering\includegraphics[width=10cm]{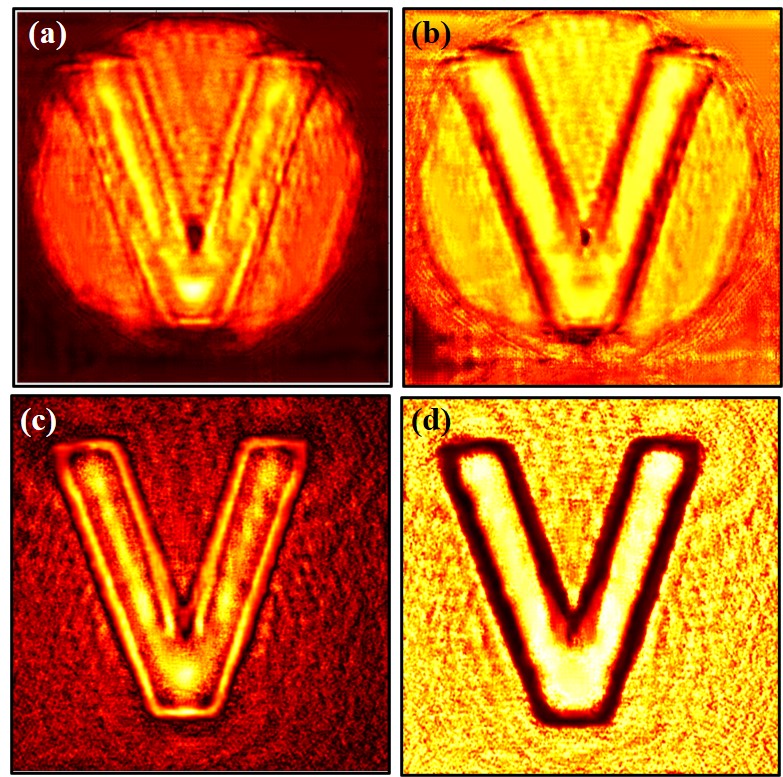}
\caption{(a-b) Conventional case results for letter V; (a)Amplitude (b)Phase, (c-d)Proposed scheme-based results for letter V; (a)Amplitude (b)Phase}
\end{figure}

\section*{Conclusion}
In conclusion, we have demonstrated a new method to record holograms in the second-order intensity correlation and the advantage of this recording appears in the quality of the reconstructed object from the hologram. In order to demonstrate the proposed scheme of recording in the holography, we recorded inline holograms in both conventional and randomness-assisted approaches and the results are compared. A twin image issue of the inline hologram is resolved with a deep learning method that uses an auto-encoder model. The quantitative features are retrieved and quality is compared in a conventional and the proposed scheme. The scheme is expected to be useful in digital holography and other imaging schemes.

\bibliography{sample}


\section*{Acknowledgements}
This work is supported by Science and Engineering Research Board (SERB) India- CORE/2019/000026. Manisha acknowledges fellowship from the IIT (BHU).

\section*{Author contributions statement}
Manisha conceived of idea and build the theoretical basis, experimental design, and completed simulation and preparation of manuscript. A.C.M. was involved in  experimental design, simulation, and preparation of the manuscript. M.R. supported in performing experiment. Z.Z provided advice and assistance, reviewing and editing the work. R.K.S was involved in supervision, formulation of research goals and aims, funding acquisition, reviewing, and editing.
\section*{Additional information}
\textbf{Competing financial interests: }The authors declare no competing financial interests. 
\end{document}